# 3D quasi-static ultrasound elastography with plane wave *in vivo*

Clement Papadacci, Ethan A. Bunting, Elisa E. Konofagou

*Abstract*— In biological tissue, an increase in elasticity is often a marker of abnormalities. Techniques such as quasi-static ultrasound elastography have been developed to assess the strain distribution in soft tissues in two dimensions using a quasi-static compression. However, as abnormalities can exhibit very heterogeneous shapes, a three dimensional approach would be necessary to accurately measure their volume and remove operator dependency. Acquisition of volumes at high rates is also critical to performing real-time imaging with a simple freehand compression. In this study, we developed for the first time a 3D quasi-static ultrasound elastography method with plane waves that estimates axial strain distribution *in vivo* in entire volumes at high volume rate. Acquisitions were performed with a 2D matrix array probe of 2.5MHz frequency and 256 elements. Plane waves were emitted at a volume rate of 100 volumes/s during a continuous motorized and freehand compression. 3D B-mode volumes and 3D cumulative axial strain volumes were successfully estimated in inclusion phantoms and in *ex vivo* canine liver before and after a high intensity focused ultrasound ablation. We also demonstrated the *in vivo* feasibility of the method using freehand compression on the calf muscle of a human volunteer and were able to retrieve 3D axial strain volume at a high volume rate depicting the differences in stiffness of the two muscles which compose the calf muscle. 3D ultrasound quasi-static elastography with plane waves could become an important technique for the imaging of the elasticity in human bodies in three dimensions using simple freehand scanning.

## I. Introduction

In biological tissues, an increase in stiffness is often a marker of abnormality [1]. Manual palpation is routinely used by clinicians to detect these variations. Over the past three decades, elastography has been developed to assess and image the mechanical properties of the muscles and organs of the human body. Originally, the term 'elastography' was introduced in the ultrasound field by Ophir et al. [2] and consisted of the estimation of local axial strains from radio-frequency (RF) cross-correlation of the ultrasound signals along the ultrasonic beam after a quasi-static compression. The idea relies on the fact that stiff tissue will deform less under compression than soft tissue, inducing a lower absolute strain on the image. This technique is better known as 'quasi-static elastography'. Elastography has since been extended to other imaging modalities such as magnetic resonance imaging [3] or optical coherence tomography [4] and in the ultrasound field, to different approaches using dynamic methods such as vibro-acoustography [5], shear wave imaging [6] or acoustic radiation force imaging [7]. The ultrasound quasi-static elastography method describes an imaging system which can be fast, non-invasive, easy to implement, conceptually very simple, and already widely used in the world of radiology thanks to several commercial implementations (Hitachi "Real-time Tissue Elastography", Siemens "eSie Touch Elastography Imaging"). In the clinic, the method enables for instance, to characterize and classify breast lesions in patients [8], [9]. However, the technique is currently limited to strain estimation in two dimensional images. This could be an issue for imaging and detecting tumors. Indeed, tumors can exhibit very heterogeneous shapes and a single cross-sectional strain map could lead to an erroneous estimation of the tumor volume. Three-dimensional strain distribution of the tumor could significantly help to accurately measure the tumor volume, in particular if the tumor is irregular. More generally, three-dimensional imaging could help to reduce intra- and inter-observer variability [10].

In addition, in most quasi-static elastographic methods, typically only the axial strain component is estimated [11] despite the fact that coupling exists between the three strain components [12]. This coupling leads to artifacts and to low signal-to-noise ratio in the axial strain distribution estimation [13]. Thus, some methods have been developed to estimate and use the lateral strain components to correct the axial strain distribution [14]–[19]. However, the strain distribution in the elevational direction remains unknown and can lead to signal decorrelation. Thus, assessing the strain distribution in the three directions would provide all components of the strain tensor, including elevational components, necessary to accurately correct axial strains [15]. Moreover, being able to fully visualize the 3D strain distributions within the tissue [20] could give access to other mechanical parameters such as the anisotropy or the elasticity values by solving the inverse problem [21].

Several different methodologies have been proposed to acquire 3D strain volumes in the existing literature. Classically,

Research reported in this publication was supported by National Institutes of Health (NIH) grants R01-EB006042 and R01-HL114358. Clement Papadacci is also supported by the Bettencourt Schueller Foundation.

C. Papadacci, E. A. Bunting and E. E. Konofagou are with the Department of Biomedical Engineering, Columbia University, New York, NY 10027 USA. (e-mail: papadacci.clement@gmail.com).

S. B. Author, Jr., was with Rice University, Houston, TX 77005 USA. He is now with the Department of Physics, Colorado State University, Fort Collins, CO 80523 USA (e-mail: author@lamar.colostate.edu).

T. C. Author is with the Electrical Engineering Department, University of Colorado, Boulder, CO 80309 USA, on leave from the National Research Institute for Metals, Tsukuba, Japan (e-mail: author@nrim.go.jp).

3D ultrasonic volumes were made from a stack of 2D ultrasonic focused images acquired by translating or rotating mechanically a 1D linear array through a stepper motor. A small motor could be integrated into the probe (wobbler probes) [22]–[24],[24] or not [25]–[29]. To acquire 3D strain volumes, the compression by the ultrasonic probe could be applied mechanically through an axial motor [22], [25]–[29], or manually (freehand compression) [24],[30], to increase clinical applicability.

Although these approaches have led to interesting results to measure breast tumors in vivo [22], or to retrieve the 3D strain components distribution in a phantom [29] or to perform freehand scanning for nearly real-time [24],[30], the need for translation of the probe mechanically was a severe limitation for real-time clinical applications to allow physicians to image 3D strains with a simple freehand scanning. Indeed, the time for the 1D array to scan the entire volume (2 seconds for the fastest case reported in literature [24]) can lead to signal decorrelation; between the first and the last 2D image acquired in the stack of 2D images contained in the 3D volume; or between the reconstructed 3D volumes before compression and after compression. This decorrelation can be due to hand motion [31] (in case of freehand compression), normal physiologic motions induced by cardiac [32] or respiratory activities [31] or to a too high strain amplitude between the two volumes [33]. Signal decorrelation is indeed a crucial issue for strain imaging as it strongly affects the signal-to-noise ratio (SNR) of the strain images [13]. Moreover, the need to minimize the acquisition time may lead to a reduction of translations and a reduction of 2D stacked images in the volume, resulting in a lower resolution in the elevation direction. In addition to a lower quality of the axial strain volume representation due to a low elevational resolution, the elevation displacement estimations become less accurate, which could be an issue to correct the axial strains [13] or to retrieve all components of the strain tensor. Finally, because the volume rate is also dependent on the number of focused emissions performed in each 2D stacked images, the number of focused emissions in transmit has also to be reduced to increase the volume rate, which implies as previously a decrease of strain estimation quality in the lateral direction.

In biomedical ultrasound, imaging at high frame rates with parallel receive beamforming has been developed to drastically increase the frame rate for various applications [34]. Typically, ultrasound images are made from multiple focused emissions, whereas high frame rate imaging (or ultrafast imaging) relies on the emission of single non-focused waves (plane waves or diverging waves) to construct the same image, enabling the frame rate to increase. Originally developed to track shear waves in biological tissue [6], high frame rate imaging has been applied to strain imaging for various applications, including cardiac strain imaging [35], electromechanical wave imaging [36], pulse wave imaging [37] or vascular strain imaging [38].

Very recently ultrafast imaging has been applied to 3D imaging using a fully programmable 2D matrix array probe for applications including cardiac imaging [39], Doppler imaging [40] and shear wave imaging [41]. It enables the acquisition of entire volumes with single transmits at a high volume rate without the need to make any compromises on the lateral or elevational resolutions.

In this study, the objective is to show the initial feasibility of performing strain estimation with plane wave imaging in three dimensions in gels and *in vivo*. For the first time we develop 3D quasi-static elastography imaging with plane waves at high volume rate using parallel receive beamforming to image axial strain distributions in entire volumes using both motorized and freehand quasi-static compression. This technique alleviated most of the decorrelation signal issues due to the very short time needed to acquire 3D volumes. With the system, we first analyzed experimentally the Point Spread Function (PSF) of the plane wave transmission compared to 256 focused emissions. Experiments were then performed in a two-layer gelatin phantom with different stiffness, a stiff cylinder and a stiff inclusion embedded in soft phantoms, and in an *ex vivo* canine liver before and after a high focused ultrasound (HIFU) ablation. Hundreds of 3D volumes were acquired at a high volume rate of 100 volumes/s during a motorized 3% continuous compression. Finally, the feasibility of the method has been shown *in vivo* on a calf of a human volunteer using freehand compression. From the Radio-Frequency (RF) data, B-mode and cumulative axial strain volumes were computed. This calculation was performed on graphic processing units (GPU) within a few seconds following acquisition. In each case, the 3D cumulative axial strain volume was able to differentiate stiff tissues from soft tissues in a full volume. 3D quasi-static elastography with plane waves could potentially constitute an important tool to image the mechanical properties of human biological tissues in 3D.

## II. Methods

### A. Ultrasound system

A fully programmable ultrasound system with 256 fully programmable channels in emission and receive (Vantage, Verasonics, Kirkland, USA) was used to control an ultrasonic 2D matrix array probe of 256 square elements (16-by-16 elements, 0.85 mm squares), with an inter-element spacing of 0.95 mm, a central frequency of 2.5MHz and a bandwith of 50% (Sonic Concepts, Bothell, USA). The volume delay-and-sum beamforming and the axial strain distribution calculations were performed on a Tesla K40 GPU (Nvidia, Santa Clara, USA). 3D rendering was computed with Amira software (Visualization Sciences Group, Burlington, USA). The calculations took a few seconds after acquisition.

### B. Experimental Setup

#### 1) Point Spread Function (PSF) analysis

Resolution and mean level were quantified using a customized resolution phantom with a 1 mm metallic bead embedded in gelatin, for a single plane wave transmission and a conventional 3D focused scheme with 16 by 16 (256) emissions focused at 30 mm depth. The phantom was scanned at 6 different depths to generate a synthetic resolution phantom. Resolution at each depth in the x- and y-directions was quantified as the -6dB width of the main lobes in the two directions. Mean level was quantified as the average of the normalized amplitude outside a sphere of 5 wavelength centered at the maximum of each PSF.

#### 2) Ex vivo study

The 2D matrix array probe was mounted on a linear motor (Fig. 1A). A square-plate compressor (10cm-by-10cm) was designed to fit the foot-print of the probe and to compress the samples with uniaxial compression. The samples were immersed inside a water tank with temperature maintained between $10^0$C and $15^0$C. A 2D real-time focused B-mode image was used to position the probe on the samples and a low axial pre-compression was applied to the samples before starting the experiments (~1 %). The samples were continuously compressed using a linear motor up to a 3% compression at a motor speed of 3% compression per second. Simultaneously, 100 plane waves were emitted from the 2D-matrix array probe using the full aperture at a rate of 100 plane waves per second in order to reconstruct 100 volumes for a total compression of 3% (Fig. 2A,B); the radio-frequency (RF) backscattered echoes from each plane wave were recorded at a frequency of 10MHz and stored in memory.

Values of Young's modulus ($E_{gelatin}$) of the manufactured phantoms (two-layer phantom and stiff inclusion) were calculated from the concentration of gelatin Bloom-275 ($C_{gelatin}$) and the formula (4) of reference [42]:

$$E_{gelatin} = 0.0034 C_{gelatin}^{2.09} \qquad (1)$$

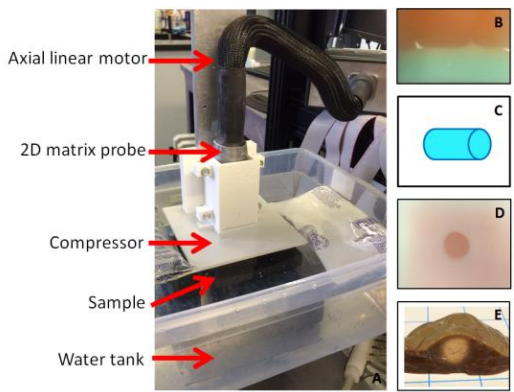

Fig. 1. Photos of the experimental setup (A) and the four ex vivo studied samples (B,C,D,E). A 2D matrix array probe was mounted on an axial linear motor. A compressor which fited the probe footprint was attached to it. The sample was positionned underneath in a cold water tank. Four different types of samples were studied: a gelatin gel with a soft layer at the top and a stiff layer at the bottom (B), a stiff cylinder embedded in a soft background (C), a stiff gelatin inclusion embedded in a soft gelatin gel (D) and an ex vivo canine liver before and after an HIFU ablation (E).

*3) Two-layer phantom*

The two-layer gelatin phantom used is shown in Fig. 1B. The first layer was made from a concentration of 3% gelatin (Bloom-275) resulting to a 4.2 kPa stiffness. The second layer was made from 12% of gelatin resulting to a 75.3 kPa stiffness. In both layers, corn stash was added to improve the backscattering properties of the gels; 1.2% in the first layer and 0.3% in the second layer to be able to see the difference on the 3D B-mode volume.

*4) Stiff cylinder in a calibrated phantom*

A commercial calibrated phantom (CIRS, customized model 049A) containing a 10.4 mm diameter cylinder of 40±8 kPa placed at 42 mm depth embedded in a 5±1 kPa background was scanned. The speckle in the cylinder and the background was identical.

*5) Stiff Inclusion*

A silicon mold of the size of the desired inclusion was created. The desired concentration of gelatin melted at a temperature of 80ºC was inserted into the mold and placed in a fridge at a temperature of 6ºC. 24 hours later, the inclusion was unmolded and embedded in the middle of a melted gelatin phantom of the desired stiffness, nearly solidified. The phantom was then placed in a fridge for another 24 hours before scanning.

A stiff gelatin inclusion (12% gelatin) with a 14.0 mm diameter with 0.3% corn stash; was thus embedded in a softer gelatin phantom (3% gelatin) with dimensions 60mmx40mmx40mm with 1.2% corn stash (Fig. 1D).

*6) Ex vivo canine liver*

The feasibility of 3D quasi-static elastography was then demonstrated in an *ex vivo* canine livers before and after a High Intensity Focused Ultrasound (HIFU) ablation (Fig. 1E). 3D quasi-static elastography was first performed on the liver without ablation. The real-time focused B-mode image of the 2D matrix array probe was used to manually position a needle in the middle of the plane-of-view. Then, the HIFU setup was positioned at the same location using the needle as a landmark. After HIFU ablation, the needle was repositioned at the same location enabling the 2D matrix array probe to return at its previous location. 30 minutes post-ablation, the 3D quasi-static elastography was performed on the ablation lesion location. The HIFU ablation was performed using a 93-element HIFU array (H-178, Sonic Concept Inc. Bothell WA, USA) generating an amplitude-modulated signal ($f_{carrier}$ = 4.5 MHz and $f_{AM}$ = 25 Hz) with an acoustic power measured in water of 5.04 W. The liver sample was sonicated for 120 s which has been shown to generate reproducible lesions similar to previous reports by our group [43].

*7) In vivo study*

The feasibility of the 3D quasi-static elastography was then demonstrated in vivo in the calf muscle of a human volunteer. The calf was set resting on a chair while the 2D matrix array probe was hand-held allowing freehand scanning. The calf muscle was continuously and smoothly compressed using the same square compressor used previously. Similarly to the *ex vivo* study, 100 2D-plane waves were emitted from the 2D-matrix array probe at a rate of 100 plane waves per second in order to reconstruct 100 volumes for the total freehand compression. The RF backscattered echoes from each plane wave were recorded at a frequency of 10MHz and stored in memory.

*C. Image formation and 3D strain calculation*

The 3D quasi-static elastography framework is depicted in Fig. 2. From RF data, a three dimensional delay-and-sum algorithm was used to beamform a 3D volume from each plane wave acquisition, resulting in a total of 100 volumes (Fig. 2B). Dynamic focusing in receive was performed with the full aperture at each point of the volumes. The lateral sizes (x and y directions) of the volumes were set to 15.2 mm corresponding to the aperture of the matrix array probe whereas the reconstructed depth was set to 30 mm-60 mm depending on the application. 64x64 lines were reconstructed with a lateral sampling of 237.5 µm in x and y directions (Fig. 2B). The axial sampling was set to 61.6 μm corresponding to a $\lambda/10$ beamforming (where $\lambda$ is the ultrasonic wavelength). From the RF beamformed signals, B-mode volumes were displayed in decibels after using a Hilbert transform and were displayed with the Amira software. The 3D incremental axial displacements between two successive volumes were estimated by normalized 1D cross-correlation [44] with a window size of 6.16 mm (corresponding to a $10\lambda$ window size as indicated in [45] and a 95% overlap) (Fig. 2C). The 3D incremental axial displacements were then integrated over time to obtain 3D cumulative axial displacements (Fig. 2D). The 3D cumulative displacements were then filtered using a 3D median filter with a pixel kernel of 3pixels-by-3pixels-by-3pixels (195μm-by-712μm-by-712μm). 3D cumulative axial strain distributions were computed from 3D cumulative displacements by applying a least-squares estimator [46] with a relatively small kernel of 646 μm (Fig. 2E). A second 3D median filter with a pixel kernel of 3pixels-by-3pixels-by-3pixels (195μm-by-712μm-by-712μm) was then applied on the strain volume. The final 3D axial strain volume was then displayed using Amira software.

Axial displacement and axial strain plots of the central line are represented. In addition, two slices in the x-z and y-z plane of the 3D axial strain volume were displayed and their locations were indicated by the orange lines on top of the 3D strain volume. Regions of interest (ROI), with the dimension of the phantoms and ablation size, were applied on the strain slices to compare the results with the known size of the phantoms and the lesions. The absolute strain values were averaged inside ($S_{in}$) and outside ($S_{out}$) the ROI with associated standard deviations ($\sigma_{in}$, $\sigma_{out}$) and an observed strain contrast (C) was calculated as follows:

$$C = 20log_{10}(\frac{S_{out}}{S_{in}}) \quad (2)$$

In the phantoms, we also estimated the strain contrast-to-noise ratio (CNR) [47] as:

$$CNR = 20log_{10}\left(\frac{2(S_{in}-S_{out})^2}{\sigma_{in}^2+\sigma_{out}^2}\right) \quad (3)$$

In addition, in the case of the *ex vivo* liver ablation a histogram presenting the strain distribution at the ablation location has been calculated before and after ablation to highlight the strain increase due to ablation stiffening. Gross pathology was used to determine the size of the lesion.

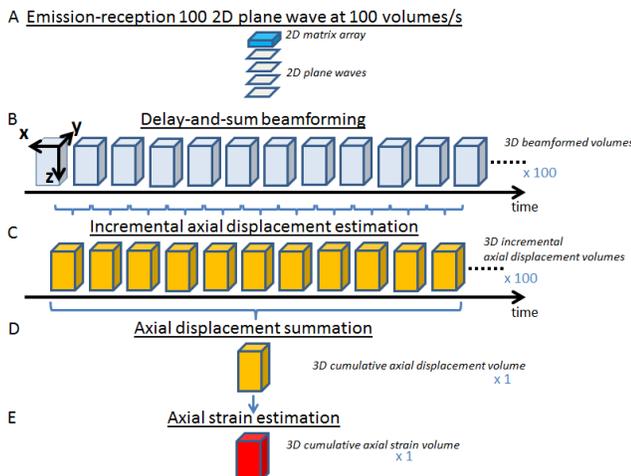

Fig. 2. Schematic of the 3D quasi-static elastography with plane waves framework. (A) The 2D matrix array emitted 100 plane waves at a rate of 100 volumes/s. (B) For each plane wave, RF backscattered echoes were stored in memories and used to calculate 3D beamformed volumes using a classical delay-and-sum beamforming. (C) 3D incremental displacement volumes were then calculated using a normalized 1D cross-correlation algorithm. (D) One 3D cumulative axial displacement volume was thus computed from the summation of the 100 incremental axial displacement volumes. (E) A 3D axial strain volume was estimated by applying a least-squares estimator and imaged using Amira Software.

III. RESULTS

A. *Point Spread Function (PSF) analysis*

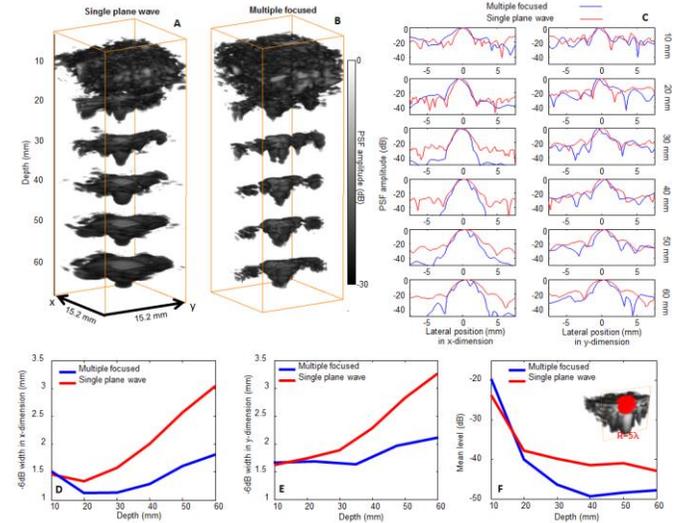

Fig. 3. Point Spread function analysis of a single plane wave transmission compared to a multiple focused scheme of 16x16 focused transmissions. 3D representation of the PSF for a single plane wave (A) and the multiple focused (B) with the associated 1D plots in x-and y-directions C. Resolution in x-and y-direction (D,E) and mean level (F) were quantified.

Figure 3 (A,B) shows the resolution phantom imaged with a single plane wave and multiple focused transmits respectively. As expected, the PSF quality is higher in the multi-focused case compared to the single plane wave in terms of resolution (D,E) and mean level (F). However, the resolution remained high for the single plane wave transmission due to the 3D focusing in receive performed in every voxel of the volume.

One can also note a quality difference in x-and y-direction due to the array manufacture Figure 3 (C).

B. *Two-layer phantom*

3D quasi-static elastography was first applied to a two-layer gelatin phantom. The layer at the top contained more scatterers. It was also stiffer (E=75.3 kPa) than the layer at the bottom (E=4.2 kPa). The 3D B-mode volume is shown in Fig. 4C. The difference in echogeneicity is clearly visible and the two layers are distinguishable. The 3D cumulative axial strain volume following a 3% compression was formed, as described in the methods. The strain difference was visible on the plot of displacements as it exhibits a different slope for the two region. This was confirmed by the strain plot (Fig. 4A,B). 3D cumulative axial strain volume is shown in Fig. 4D wih the associated slices in x-z plane (Fig. 4E) and in x-y plane (Fig. 4F). The softer layer exhibited in average a higher absolute strain ($S_{in}$=2.2%±0.2%) than the stiffer layer ($S_{out}$=0.4%±0.2%), as it was expected. The observed strain contrast and CNR were calculated and gave values of $C = 14.8$ dB and CNR=38.2 dB respectively. The boundary between the two layers on the 3D strain volume was in agreement with the boundary found on the 3D B-mode volume.

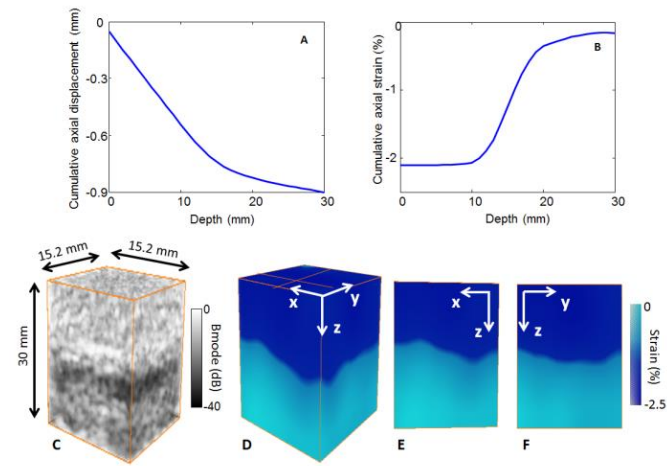

Fig. 4.  Cumulative axial displacement and strain plots of the central line of the strain volume (A,B). 3D B-mode volume (C) and 3D cumulative axial strain volume (D) of the two-layer phantom. The layers were distinguishable on the Bmode volume due to the difference in scatterers concentration (C). The layers were also distinguishable on the 3D strain volume (D), absolute strain was high for the soft layer whereas it was low for the stiff layer. Slices of the 3D strain volume were displayed in x-z plane (E) and in x-y plane (F).

### C. Stiff cylinder

3D quasi-static elastography was then applied to a a stiff cylinder (40±8 kPa) embedded in a soft calibrated phantom (5±1 kPa). The cylinder could not be seen on the 3D B-mode volume (Fig. 5C) due to the identical concentration of scatterer inside and outside the cylinder. However, the cylinder could be detected on the axial displacement curve (Fig. 5A, inflection between 35mm and 45mm) and the axial strain curve (Fig. 5B). The 3D axial strain volume (Fig. 5D) enabled to visualize the cylinder shape. The lower absolute strain ($S_{in}$=-0.6%±0.3%) indicated that the cylinder was stiffer than the surroundings ($S_{out}$=3.2%±0.6%). Fig. 5E and Fig. 5F show the cross section of the 3D strain volume in x-z plane and y-z plane respectively with a circle of the size of the cylinder diameter. The observed contrast and CNR were calculated and found to be equal to: $C$ = 14.5 dB and CNR=29.5 dB respectively.

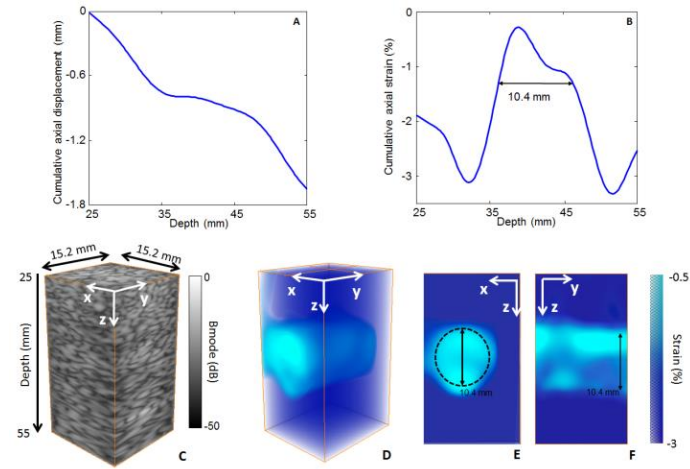

Fig. 5.  Cumulative axial displacement and strain plots of the central line of the strain volume (A,B) for the stiff cylinder. 3D B-mode volume (C), 3D axial strain volume (D) and slices of the strain volume (E,F) of the stiff cylinder embedded in the soft phantom. The cylinder was not distinguishable on the B-mode volume due to the similar concentration of scatterers (C). However, the cylinder was distinguishable on the 3D axial strain volume (D) as the absolute axial strain was low in the stiff ciliner whereas it was high in the soft phantom. Slices of the 3D axial strain volume are displayed in x-z plane (E) and in x-y plane (F) with the ROI. Dashed contour delineates the cylinder.

### D. Stiff inclusion

The method was then applied to a stiff gelatin inclusion embedded in the softer gelatin phantom. The results are displayed in Fig. 6. The inclusion was clearly identifiable on the 3D B-mode volume as it was made with a smaller concentration of scatterer (Fig. 6C). The 14 mm stiff inclusion was also identifiable on the axial displacement plot (Fig. 6A), the axial strain plot (Fig. 6B), and the 3D axial strain volume (Fig. 6D), due to the lower absolute strain estimated at the inclusion location ($S_{in}$=0.3%±0.4%) compared to the surrounding ($S_{out}$=1.5%±0.4%). Fig. 6E and Fig. 6F show the cross-section of the 3D axial strain volume in the x-z plane and the y-z plane respectively with a 14 mm diameter circle depicting the ROI. The observed contrast $C$ = 13.9 dB and CNR = 25.6 dB, were lower than the two-layer and stiff cylinder phantom cases. This can be explained by the geometry of the inclusion. Indeed, a lower contrast transfer efficiency [48], which represents the ratio of elasticity contrast measured from strain estimation to the true contrast, was expected in the case of a spherical inclusion [49]. In addition, we can observe some of the predicted shadowing artifacts outside the inclusions on Fig. 6E and Fig. 6F.

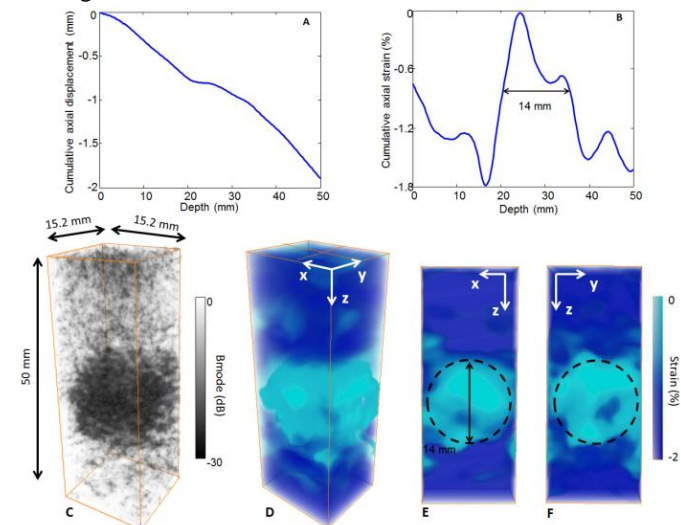

Fig. 6.  Cumulative axial displacement and strain plots of the central line of the strain volume (A,B). 3D B-mode volume (C), 3D axial strain volume (D) and slices of the strain volume (E,F) of the stiff inclusion embedded in the soft phantom. The inclusion was distinguishable on the B-mode volume due to the different concentration of scatterers (C). The inclusion was also distinguishable on the 3D axial strain volume (D), absolute axial strain was low for the stiff inclusion whereas it was high for the soft phantom. Slices of the 3D strain volume were displayed in x-z plane (E) and in x-y plane (F). Dashed contour delineates the inclusion.

### E. Ex vivo liver

3D quasi-static elastography was obtained to an *ex vivo* canine liver before and after a HIFU ablation. The results are shown in Fig. 7 and Fig. 8. Before ablation, the 3D B-mode volume exhibited a heterogeneous echogenicity (Fig. 7A). This heterogeneity is also noticeable on the 3D axial strain volume (Fig. 7B) and 2D strain images (Fig. 7C and D), revealing the structural complexity of the liver. After ablation, the 3D B-mode volume was more echogenic at the ablation location. However, only based on the 3D B-mode volume, it was not possible to precisely detect the ablation lesion. On the 3D axial strain volume (Fig. 7F), one can notice a decrease on the absolute strain at the HIFU ablation location as expected. It was also visible on the axial displacement and strain plots (Fig. 7I,J). Indeed, the effect of HIFU ablation in biological tissue is an increase in stiffness [43]. According to the average strain inside ($S_{in}$=0.25%±1.1%) and outside the ablation lesion ($S_{out}$=0.8%±1.3%) the observed contrast $C = 10.1$ dB was calculated and was found comparable to the stiff inclusion. However, the artifacts were higher in this case as it is showed by the higher standard deviation from the average strain calculation. These artifacts could be due to the more complex structure of the liver compared to the phantoms.

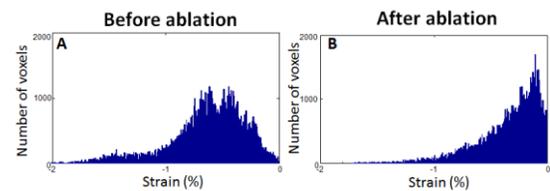

Fig. 8. Strain distribution at the ablation location before ablation (A) and after ablation (B). Absolute strain is lower after ablation due to stiffening.

### F. In vivo calf of a human volunteer

The feasability of method was then demonstrated *in vivo* on the upper calf of a human volunteer. From the acquisition, we were able to reconstruct the 3D B-mode volumes (Fig. 9C) and slices (Fig. 9D and E). From the freehand compression and the transmission of 100 plane waves at 100 volumes/s, we were able to construct a 3D axial strain volume (Fig. 9F). Axial displacement and strain on the central line were also displayed (Fig. 9A,B). We could detect the two muscles composing the calf muscle, the gastrocnemius muscle at the top and the soleus muscle at the bottom which exhibited similar strain values. The softer layer in between the two muscles was consistent with the location of the fascia layer that embeds the muscles (i.e Fig. 9F). This findings are in a good agreement with [50] which found similar results with quasi-static elastography in two dimensions with a linear array probe.

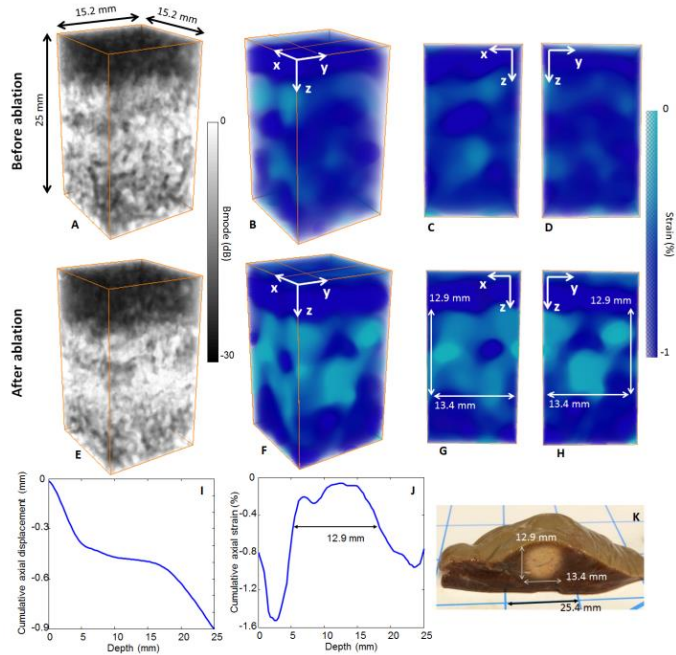

Fig. 7. 3D B-mode volumes (A,E), 3D strain volumes (B,F) and slices of the strain volumes (C,D,G,H) of an *ex vivo* canine liver before (A,B,C,D) and after an HIFU ablation (E,F,G,H). Before ablation, the 3D B-mode volume displays scattering differences due to the complex structure of the liver (A). The 3D strain volume (B) and the slices (C,D) also show heterogeneities in elastic properties with strain variations. After ablation, the brightness of the 3D B-mode increases (E), and the 3D strain volume (F) and slices (G,H) reveal the stiffer ablation lesion. The cumulative axial displacement (I) and strain (J) were displayed at the central line after the ablation. Gross pathology (K) was in a good agreement with the lesion measurements on the strain volume.

Fig. 8 displays the axial strain distribution over the lesion volume before (Fig. 8A) and after (Fig. 8B) the HIFU ablation. Decrease of the absolute strain after ablation confirmed the liver stiffening.

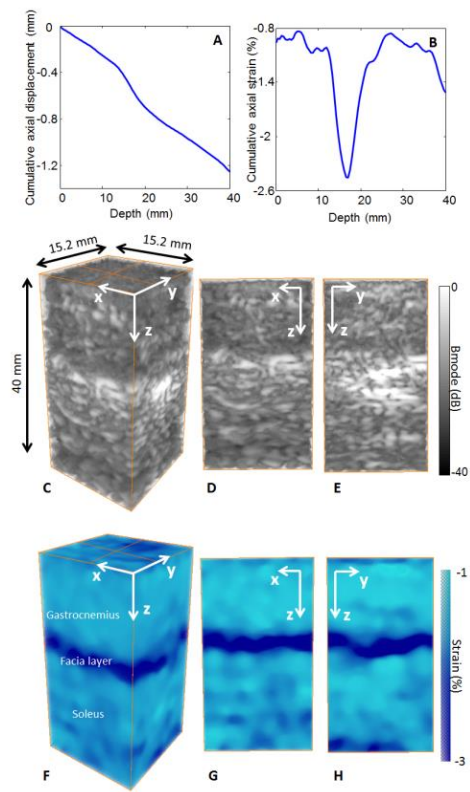

Fig. 9. Cumulative axial displacement and strain plots of the central line of the strain volume (A,B). 3D B-mode volume (C) and slices (D,E) and 3D strain volume (F) and slices (G,H) of an *in vivo* calf muscle of a human volunteer. The muscles of the gastrocnemius at the top and the soleus at bottom were identified and exhibited a similar straiun value. The facia layer was identified at their boundary and found to have a higher absolute strain.

## IV. Discussion

In this study, the objectives were to develop a new 3D axial strain method using plane waves at high volume rate. We demonstrated the feasibility of 3D quasi-static elastography based on three dimensional quasi-static elastography coupled with plane wave imaging at a high volume rate using a 2D matrix array probe. We successfully estimated axial strains in three dimensions in phantoms, in an *ex vivo* biological tissues for ablation lesion detection, and *in vivo* in the calf muscle of a human volunteer.

We first compared experimentally with the same system the PSF of a single plane wave transmission with a multiple focused scheme. We combined a 3% motorized continuous quasi-static compression with 3D strain imaging with plane waves, in order to acquire 100 volumes at a volume rate of 100 volumes/s. This very high volume rate allowed us to compute one 3D volume of cumulative axial strains made from 100 incremental axial displacements.

The method was validated in phantoms and showed good contrast. We were able to detect strain differences on a two-layered phantom composed of different stiffness. In addition, a 10.4 mm diameter stiff cylinder and a 14 mm diameter stiff inclusion embedded in a soft background were successfully detected and visualized. A validation of the size of the phantoms compared to the measured strains will be performed in a future study. We also showed the feasibility of the method in an *ex vivo* canine liver before and after an HIFU ablation by successfully detecting the stiffer lesion after HIFU ablation. Finally, the feasibility of performing 3D quasi-static elastography with plane waves at a high volume rate *in vivo* was demonstrated on the calf muscle of a human volunteer by assessing 3D axial strain volume with a simple freehand compression, in which the composition of the muscle was identified and in good agreement with literature [50].

Using only one transmit plane wave to construct an entire volume enabled high volume rates and eliminated signal decorrelation occurring when multiple acquisitions from mechanical translation of a 1D array were required to construct the volume. The acquisition time posed thus no longer a limitation.

For *in vivo* applications, such as breast cancer detection, the high volume rate used in this study will likely reduce signal decorrelations from physiologic motion such as the respiration or the heart rhythm. It also reduced artifacts from hand motions in the case of freehand scanning.

In addition, the high volume rate enabled high temporal resolution which leaded to lower strain amplitude between consecutive volumes. In this case, the Strain Filter framework [33] predicted an improvement in the strain estimator performance by reducing signal decorrelation and increasing the correlation coefficient. In this study a volume rate of 100 volumes/s was used but the method enables the volume rate to be increased up to 3000 volumes/s, which could be used to optimize the technique based on the Strain Filter. In clinical practice, the compression motion applied can be different among operators, the high volume rate could be used to ensure the same inter-volume displacements maintaining the same strain estimation quality.

In addition, the high number of volumes acquired enabled the calculation of numerous incremental axial displacement volumes in order to calculate, in this study, one volume of 3D cumulative axial displacement and one 3D cumulative axial strain volume. This is a similar approach to the technique of multicompression averaging, which has been shown to increase strain signal-to-noise ratio by decreasing random noise [51].

This method based on axial strain estimation could also be extended to lateral and elevational strains to correct the axial strain [20] or to obtain the full strain tensor in biological tissues, with a high temporal resolution. This constitutes ongoing work by our group.

Due to the associated high volume rates, this method could also be extended to moving organs such as the heart. Myocardial elastography [52], a method for the estimation of the strain distribution in the heart, could be measured in entire volumes at high temporal resolution.

The implementation on GPUs enabled us to obtain the 3D B-mode and the 3D axial strain in a few seconds following the acquisitions. The reconstruction time depends on the depth, the lateral and axial sampling. For a reconstruction at 5 cm depth sampled at $\lambda/10$ with 64x64 lines, 8.5s were needed. This method has thus the potential to be implemented in real time by optimizing the processing.

The objective of this study was to show the initial feasibility of estimating strain with plane wave in three dimensions and demonstrating that, even in the presence of poorer sonographic image quality in terms of SNR and spatial resolution, strain can be imaged to reveal structures of distinct elasticity.

The study presented has several limitations. The main limitation of this study is the poor image quality. However, we found with the PSF study that even with only one plane wave transmission, the quality of the PSF remained high compared to the focused transmissions. The resolution decreases by only 25% compared to focusing transmits at 30 mm depth. Even if the mean level of the plane wave transmission was found to be higher (10-15dB) than the focusing transmits, this loss, which can be related to the contrast of the B-mode image, might be not as critical for displacement estimation. Indeed, few studies reported very low contrast in biological tissue, such as the heart imaged with a single transmit, but still very high displacement and strain estimations [36]. The strain estimation quality from focusing transmits compared to a single transmission was also studied in terms of expected SNR in [35] where very similar results were found.

The PSF study indicates that the reduced image quality is more likely due to the low frequency of the probe (2.5MHz), the small aperture, the low number of elements in both directions (16x16 elements), the low frequency bandwith (50%) of the probe used, which resulted in a low signal-to-noise ratio in the strain estimation [33] and the contrast transfer efficiency [48]. This limitation could be overcome by using a 1024 elements probe with a larger bandwidth, controlled by two synchronized ultrasound systems with 256 fully programmable channels combined with synthetic emissions as it is performed in [39]. The quality of the estimation could also be increased by increasing resolution, by implementing spatial coherent compounding in three dimensions from the transmission of multiple tilted plane waves [39] instead of using only one plane wave performed in this study. Another limitation is the small field of view induced by plane wave transmission in the lateral and elevation direction from the small aperture of the probe. This issue could be addressed by implementing diverging wave transmits instead of plane waves [39] while keeping a high volume rate. These optimization strategies go beyond the scope of this paper but constitute topics of ongoing research in our group.

## V. CONCLUSION

We successfully developed and implemented 3D quasi-static elastography with plane waves at high volume rate. By transmitting 100 plane waves at a volume rate of 100 volumes/s, we were able to detect and image the axial strain distribution in a two-layer gelatin phantom of two different stiffness, a stiff cylinder and a stiff inclusion embedded in a soft phantom throughout the entire volume in three dimensions. Moreover, we estimated the 3D axial strain in an *ex vivo* canine liver before and after HIFU ablation and were able to map the axial strain distribution in three dimensions to detect the ablation lesion. Finally, we demonstrated *in vivo* feasibility of 3D quasi-static elastography with plane waves on the calf of a human volunteer with a simple freehand scanning at a high volume rates. Due to the high volume rate, 3D quasi-static elastography for *in vivo* applications could have a major impact for three dimensional elasticity imaging.

## ACKNOWLEDGMENTS

Research reported in this publication was supported by National Institutes of Health (NIH) grants R01-EB006042 and R01-HL114358. Clement Papadacci is also supported by the Bettencourt Schueller Foundation. The authors would like to thank Thomas Payen Ph.D and Pablo Abreu for the additional help on this project.